\documentstyle[preprint,aps,epsf,floats,tighten]{revtex}
\input{epsf}
\begin{document}
\draft
\preprint{
\vbox{\hbox{JHU--TIPAC--99006}
      \hbox{UTPT-99-07}
      \hbox{June 1, 1999} }}
\title{Charmonium production at neutrino factories}
\author{\normalsize Alexey A. Petrov\thanks{petrov@pha.jhu.edu}}
\address{Department of Physics and Astronomy, Johns Hopkins University,
  Baltimore, MD 21218, USA}
\author{Tibor Torma\thanks{kakukk@physics.utoronto.ca}}
\address{Department of Physics, University of Toronto, Toronto, Ontario, Canada M5S~1A7}
\date{\today}
\maketitle
\begin{abstract}
At existing and planned neutrino factories (high energy and high intensity
neutrino beam facilities) precision studies of QCD in neutrino-nucleon interactions
are a realistic opportunity.
We investigate charmonium production in fixed target neutrino
experiments. We find that $J/\psi$ production in neutrino-nucleon collision
is dominated by the color octet $^3S_1$ NRQCD matrix element in
a neutral current process, which is not accessible in photo or leptoproduction.
Neutrino experiments at a future Muon 
Collider will acquire sufficient event rate to accurately measure color octet
matrix element contributions. 
The currently running high energy neutrino experiments NOMAD and NuTeV
could also observe several such events.
\end{abstract}

\narrowtext

\section{Introduction}

One of the most promising future high energy facilities is the recently proposed Muon Collider. 
In order to facilitate a decision on whether and how it should be built, all the
various uses it can be put to should be assessed. The highly collimated and intense 
neutrino beams unavoidably generated by muon decay provide a unique opportunity for 
precision studies of QCD and electroweak physics. An excellent example of such 
investigations is related to the ongoing issue of the validity of the 
Nonrelativistic QCD (NRQCD) expansion for
charmonium states and the extraction of the so-called color octet matrix elements.
The fact that heavy quarkonium represents a non-relativistic quantum-mechanical system
significantly simplifies its theoretical studies. In particular, the presence of several
important scales in the quarkonium system, $M$, $Mv$ and $Mv^2$ ($\approx\Lambda_{QCD}$) 
where $v$ is a small parameter (relative velocity of quarks in the quarkonium state)
allows separation of physical effects occurring at different scales~\cite{NRQCDrev}.

A large excess of prompt $J/\psi$'s and $\psi^\prime$'s at the Tevatron
over to the predictions of the color singlet model, i.e. the model which
postulates that only quarks in a relative color singlet state can evolve into a
charmonium, sparked both experimental and theoretical interest and resulted in 
the realization of the importance of contributions generated by the operators
involving quark states in a relative color octet configuration. The emerging effective 
theory (NRQCD), systematically describing these processes, factorizes the charmonium 
production cross section in the form
\begin{equation} \label{factor}
\sigma(A + B \to H +X) = \sum_n \frac{F_n}{m_c^{d_n-4}}\,
\langle 0 | {\cal O}^H_n | 0 \rangle ,
\end{equation}
where $F_n$ are short-distance coefficients containing the perturbatively
calculable hard physics of the production of a $[c\overline{c}]$ system at almost zero
relative velocity (like $\gamma g \to c \bar c g ,~ q \bar q  \to c \bar c g$,
etc.), expressed as a series in $\alpha_s(m_c)$. Here, the index $n$ incorporates a spectral
decomposition of the quarkonium state in terms of the quantum numbers $\,^{2S+1}L_J^{(color)}$ 
of the $[c\bar c]$ system, as well as the number of additional derivatives acting on the
heavy quark fields. The essence of NRQCD is to organize the above expansion in powers of
the heavy quark velocity $v$ within the hadron, and it can be further generalized to include
other heavy quarkonium-like systems, such as heavy hybrids~\cite{Bc,Hh}.
Eq.~(\ref{factor}) puts all nonperturbative long-distance information into the NRQCD 
matrix elements, which describe the evolution of the $[c\overline{c}]$ system into 
a charmonium plus soft hadrons -- a process that cannot be calculated at present from 
first principles.

Several attempts have been made to determine these NRQCD matrix elements from various experiments.
The processes involved are sensitive to various linear combinations of NRQCD matrix elements.
The problem is aggravated by the usually very large theoretical uncertainties involved in these
calculations (on the order of $50-100\%$), due to higher twist effects, uncalculated
and/or incalculable higher order perturbative and nonperturbative contributions. In this
situation any independent determination of these quantities should be welcome.

A major advantage of using the neutrino beam is that, at leading order in $\alpha_s$,
the spin structure of the $\nu Z$ coupling selects a certain combination of octet operators.
The largest contribution is from the one with the quantum numbers~$\,^3S_1^{(8)}$. Of course,
order of magnitude measurements of the size of the matrix elements of this operator have already
been performed for the $J/\psi$ and $\psi^\prime$, as well as for the $\chi_{cJ}$ states. The
estimates of these matrix elements mostly come from Tevatron fits to hadroproduction cross sections
for the $J/\psi$ and $\chi_{cJ}$ and yield, with large theoretical errors~\cite{TeV1},
\begin{eqnarray} \label{estim}
&&\langle 0 | {\cal O}_8^{J/\psi}(^3S_1) | 0 \rangle \sim 0.01~\rm{GeV}^3,~~~
\mbox{and}
\nonumber\\
&&\langle 0 | {\cal O}_8^{\chi_{c0}}(^3S_1) | 0 \rangle \sim 0.01~\rm{GeV}^3.
\end{eqnarray}
These values are consistent, within a $\sim50\%$ accuracy level, with the value
found from $Z$ decay at LEP~\cite{leiblep} (the latter does not separate cascade
and direct production, so the value of $0.019\,GeV^3$ is understandably larger than the
one in Eq.~(\ref{estim})).

There are, however, large discrepancies between the Tevatron fits and the values of
$\chi_{cJ}$ matrix elements obtained from B decays~\cite{B}, and between various determinations of
$\langle 0 |{\cal O}^{J/\psi}_8(^3S_1) | 0 \rangle$ from the Tevatron fits. Clearly, new results
from HERA leptoproduction experiments would not clarify the situation as at leading order 
the process $\gamma^* g \to [c \bar c]_8(^3S_1)$ is forbidden by parity conservation of
strong interactions. In this situation other determinations are welcome and desired.

The present paper is an exploratory investigation of the main features of inclusive charmonium
production in $\nu N$ collisions. This process parallels $J/\psi$ leptoproduction,
in which case Fleming and Mehen~\cite{leptoprod} found that the ${\cal O}(\alpha_s^2)$
contribution to the total $e\,N\to e\,J/\psi\,X$ cross section is small compared to the
color octet ${\cal O}(\alpha_s)$ contribution. A set of cuts, requiring an energetic
gluon jet well separated from the $J/\psi$, enhances the ${\cal O}(\alpha_s^2)$ contributions,
but then the color singlet contribution will dominate. These cuts, however, leave behind
only a small part of the total cross section. We don't expect that either the difference in
the spin structure or the $\left(m_W/m_Z\right)^4\approx0.6$ suppression of neutral 
current (NC) versus charged current (CC) events
can change this picture, so that we feel justified to calculate only the
${\cal O}(\alpha_s)$ contributions. We will find, however, that while the leptoproduction
of $J/\psi$ is not sensitive to the $\,^3S_1^{(8)}$ matrix element, and measures one combination
of $\,^1S_0^{(8)}$ and $\,^3P_J^{(8)}$, measuring the $Q^2$ distribution in our process allows
for a determination of both the $\,^3S_1^{(8)}$ and the $\,^3P_J^{(8)}$ matrix elements.
The difference is due to a difference in the spin structure of the $Z$ and photon couplings.

The relative size of the $\,^3S_1^{(8)}$ and the $\,^3P_J^{(8)}$ contributions to the differential
cross sections turns out to change drastically around $Q^2\sim10\,GeV^2$. This fact and the
less steep decrease of the differential cross section (compared to leptoproduction) together
allow an easy separation of the two contributing matrix elements.

We will find that the hard process at small $Q^2$ favors the $^3S_1^{(8)}$ contribution 
by a significant factor. In an experiment which cannot easily separate direct $J/\psi$
production from a $\chi_{cJ}$ cascading down to a $J/\psi$ (such is the situation at
LEP, for example), these cascades enhance the observed cross section by a significant
amount. In particular, given the estimates of Eq.~(\ref{estim}), we find
$\sigma (\nu N \to \chi_{c2} + X) \sim 5\, \sigma (\nu N \to J/\psi + X)\,$! Noting that
$BR (\chi_{c2} \to J/\psi \gamma) = 13.5 \%$ we conclude that $J/\psi$ production
via this cascade mechanism is of the same order of magnitude as via the direct route.
It is however easy to include the effect of these unresolved cascades into the formalism
by simply replacing the actual $\langle 0 | {\cal O}^{J/\psi}_8(^3S_1) | 0 \rangle$
matrix element with a $\langle 0 | \widehat{\cal O}^{J/\psi}_8(^3S_1) | 0 \rangle$
which takes into account this cascade factor~\cite{leiblep}. The rate of $\chi_{cJ}$
production can then easily be related to our $\,^3S_1^{(8)}$ contribution to
the $J/\psi$ production rate, because in that case there are no other competing matrix elements.

The large number of events expected at the muon collider will certainly allow to study 
the entire spectrum of charmonium states. Due to the fact that the dominant $\,^1S_0^{(8)}$ 
matrix element of the $h_c$ and the dominant
$\,^3S_1^{(8)}$ of the $\chi_{cJ}$ are related by heavy quark symmetry, our results can be trivially
translated to a prediction of these rates. The only relevant matrix element above is measurable
in any other cascade $J/\psi$ production process.

In the present study we also discuss what we can learn about these matrix elements at existing
neutrino facilities. As we show below, the mere event of the detection of charmonium states in
current experiments (NOMAD or NuTeV) would imply the presence of the color-octet structures
predicted by the NRQCD factorization formalism. The main question to be addressed here is the
smallness of the event number. Since the detectors used in  current neutrino experiments are
optimized for the observation of neutrino-related  phenomena (such as neutrino oscillations)
and not for charmonium detection, we shall concentrate on $J/\psi$ production, which has the
cleanest experimental signature of all charmonium states. However cascade processes should also
be included, helping to increase the event number. We will find that both at NOMAD and NuTeV, 
the event rate is on the verge of observability.

The rest of the paper is organized as follows. In Sec.~\ref{sec:calc} we explain the velocity scaling
of the NRQCD matrix elements that we use and provide analytical formulas for the structure
functions~$F_n$. We discuss the numerical results for the Muon Collider, NOMAD and NuTeV in detail
in Sec.~\ref{sec:numbers} and offer concluding remarks in the Conclusions.

\section{The calculation}\label{sec:calc}

\begin{figure}[htb]
\begin{center}
\mbox{\epsfxsize=6in\epsfbox{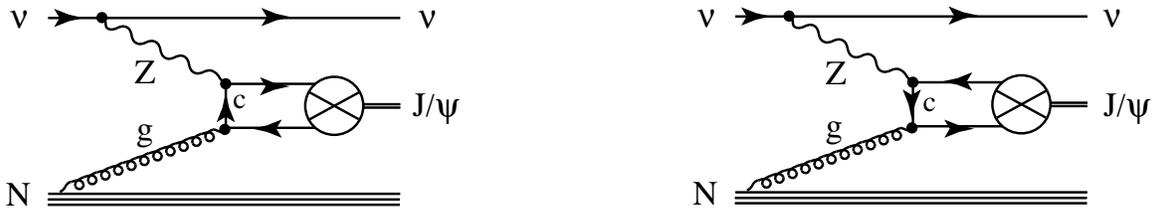}}
\end{center}
\caption{\label{fig:leadinggraphs}The leading ${\cal O}(\alpha_s)$ graphs.
Note the absence of any $W$ exchange graphs.}
\end{figure}

At leading order, ${\cal O}(\alpha_s)$, only neutral current $gZ$ interactions contribute
to charmonium production and Fig.~\ref{fig:leadinggraphs} shows the relevant Feynman
diagrams. At ${\cal O}(\alpha_s^2)$, in addition to the color singlet contribution,
one may expect some enhancement from charged current
interactions due to the difference in the couplings and the propagators, but we do not expect
that this could drastically change the cross section to invalidate our order-of-magnitude estimates.

Before proceeding with the calculation, let us briefly review the velocity counting rules 
of the charmonium production matrix elements which will be used below. The charmonium 
production matrix elements are defined in NRQCD as
\begin{equation}
\langle 0 | {\cal O}^H_n | 0 \rangle \,=\,
\sum_X \sum_{m_J}\,
\langle 0 | {\cal K}_n | H_{m_J} + X \rangle\,
\langle H_{m_J} + X | {\cal K}^\prime_n | 0 \rangle
\end{equation}
where the operators ${\cal K}^{(\prime)}_n$ are bilinear in the heavy quark
fields. ($m_J$ is the charmonium spin $z$-component.) Each matrix element is 
proportional to a power of the relative velocity~$v$
of the heavy quarks. Each spatial derivative acting on a heavy quark field introduces one power
of~$v$. The nonperturbative evolution of the $c\overline{c}$ system proceeds through multipole
radiation of soft gluons, which introduce additional powers of the velocity. For the
$J/\psi$ and $\psi^\prime$, the leading matrix element is at $m_c^3v^3$,
\begin{equation}
\langle 0 | {\cal O}^{J/\psi}_1(^3S_1) | 0 \rangle \,=\,
\langle 0 | \chi^\dagger {\bf \sigma} \psi | J/\psi + X \rangle\,
\langle J/\psi + X | \psi^\dagger {\bf \sigma} \chi | 0 \rangle,
\end{equation}
which clearly selects the dominant Fock state. We observe, however, that this
matrix element corresponds to a $c\overline{c}$ system in a color singlet state, which
can be produced only at subleading order, ${\cal O}(\alpha_s^{2})$. 
Therefore, its contribution will be smaller than that of 
the velocity-subleading color octet configuration, such as
\widetext
\begin{eqnarray}
\langle 0 | {\cal O}^{J/\psi}_8(^3S_1) | 0 \rangle \,&=&\,
\langle 0 | \chi^\dagger {\bf \sigma} T^a \psi | J/\psi + X \rangle\,
\langle J/\psi + X | \psi^\dagger {\bf \sigma} T^a \chi | 0 \rangle,
\nonumber \\
\langle 0 | {\cal O}^{J/\psi}_8(^1S_0) | 0 \rangle \,&=&\,
\langle 0 | \chi^\dagger T^a \psi | J/\psi + X \rangle\,
\langle J/\psi + X | \psi^\dagger T^a \chi | 0 \rangle,
\\
\langle 0 | {\cal O}^{J/\psi}_8(^3P_J) | 0 \rangle &=&
\langle 0 | \chi^\dagger \left[
-\frac{i}{2} { D^{\{i} \sigma^{k \} }} \right] T^a \psi | J/\psi + X \rangle\,
\langle J/\psi + X | \psi^\dagger \left[
-\frac{i}{2} { D^{ \{ i} \sigma^{k \}}} \right] T^a
\chi | 0 \rangle,
\nonumber \\
\langle 0 | {\cal O}^{J/\psi}_8(^1P_0) | 0 \rangle &=&
\langle 0 | \chi^\dagger \left[
-\frac{i}{2}\, {\bf D} \right] T^a \psi |J/\psi  + X \rangle\,
\langle J/\psi + X | \psi^\dagger \left[
-\frac{i}{2} {\bf D} \right] T^a
\chi | 0 \rangle,
\nonumber 
\end{eqnarray}
where $\{...\}$ represents scalar, vector or traceless symmetric tensor 
contraction of indices. Five of these configurations, namely those with 
the quantum numbers $\,^1S_0^{(8)}$, $\,^3S_1^{(8)}$, $\,^3P_J^{(8)}$ for $J=0,1,2$, 
scale as $v^7$. The $\,^1P_0^{(8)}$ matrix element is negligible, of the 
order ${\cal O}(v^{11})$, and will be neglected hereafter.

In the case of the p-wave states $\chi_{cJ}$ we encounter a similar situation.
The leading matrix elements are again suppressed by one factor of $\alpha_s$,
and are color singlets at $m_c^5v^5$. They project out the
dominant p-wave $c \bar c$ combination
\begin{eqnarray}
\langle 0 | {\cal O}^{\chi_{c0}}_1(^3P_0) | 0 \rangle &=&
\frac{1}{3}\, \langle 0 | \chi^\dagger \left[
-\frac{i}{2}\, {\bf D \cdot \sigma} \right] \psi |\chi_{c0}  + X \rangle\,
\langle \chi_{c0} + X | \psi^\dagger \left[
-\frac{i}{2} {\bf D \cdot \sigma} \right]
\chi | 0 \rangle
\nonumber \\
\langle 0 | {\cal O}^{\chi_{c1}}_1(^3P_1) | 0 \rangle &=&
\frac{1}{2} \langle 0 | \chi^\dagger \left[
-\frac{i}{2} {\bf D \times \sigma} \right] \psi |\chi_{c1}  + X \rangle\,
\langle \chi_{c1} + X | \psi^\dagger \left[
-\frac{i}{2} {\bf D \times \sigma} \right]
\chi | 0 \rangle
\\
\langle 0 | {\cal O}^{\chi_{c2}}_1(^3P_2) | 0 \rangle &=&
\langle 0 | \chi^\dagger \left[
-\frac{i}{2} { D^{(i} \sigma^{k)}} \right] \psi |\chi_{c2}  + X \rangle\,
\langle \chi_{c2} + X | \psi^\dagger \left[
-\frac{i}{2} { D^{(i} \sigma^{k)}} \right]
\chi | 0 \rangle.
\nonumber
\end{eqnarray}
The leading matrix elements in the velocity expansion are $\,^3P_J^{(1)}\sim m_c^5v^5$ and
$\,^3S_1^{(8)}\sim m_c^3v^5$, but the $\,^3S_1^{(8)}$ configuration is produced
at leading order in $\alpha_s$:
\begin{equation}
\langle 0 | {\cal O}^{\chi_{cJ}}_8(^3S_1) | 0 \rangle =
\langle 0 | \chi^\dagger {\bf \sigma} T^a \psi |  {\chi_{cJ}} + X
\rangle
\langle {\chi_{cJ}} + X | \psi^\dagger {\bf \sigma} T^a \chi | 0 \rangle.
\end{equation}

The velocity counting rules in NRQCD require, for the most interesting case of
the~$J/\psi$, to calculate the contributions of the ${\cal O}(v^7)$ matrix elements,
$\,^1S_0^{(8)},\,^3S_1^{(8)},\,^3P_J^{(8)}$ (heavy quark symmetry requires
$\,^3P_J^{(8)}=\mbox{$(2J+1)\,^3P_0^{(8)}$}$). We note that the same structure functions
can be used to evaluate all contributions to $\chi_c$, for which $\,^3S_1^{(8)}$ is leading,
of order ${\cal O}(v^5)$, and to $h_c$, for which $\,^1S_0^{(8)}$ is leading, also of
order ${\cal O}(v^5)$.
We do not calculate the hard-to-identify $\eta_c$, where the
uncalculated $ \,^1P_0^{(8)}$ would also contribute at leading order (this matrix element
is completely unimportant for the $J/\psi$, as it is suppressed as ${\cal O}(v^{11})$ in that case).
We also note that parity and $J$~conservation forbids all interference terms in Eq.~(\ref{eq:struct1}).

The calculation is quite straightforward and we quote the result in the form of the following
structure functions~$h_n\left(Q^2\right)$:
\begin{equation}
\frac{d\sigma\left(s,Q^2\right)}{dQ^2}=
\frac{\pi^2\alpha^2\alpha_s}{3\sin^42\theta_W}\ \frac{1}{\,
     {{\left( {Q^2} +
          {{{m_Z}}^2} \right)
         }^2}}
\times\sum_n\frac{\langle 0|{\cal O}_n|0\rangle}{m_c^3}
\int_{\frac{Q^2+4m_c^2}{s}}^1dx\,f_{g/N}\left(x,Q^2\right) h_n\left(y,Q^2\right),
\label{eq:struct1}
\end{equation}
where $s$ is the total invariant mass of the $\nu N$ system, $x$ is
the momentum fraction of the incoming gluon and~$-Q^2$ is the momentum-squared
transferred from the leptonic system; $y=\frac{Q^2+4m^2}{sx}$. Because we are
doing a tree level calculation, $\alpha_s$~does not run; in the actual calculation
we will choose the educated guess of evaluating $\alpha_s\Rightarrow\alpha_s\left(Q^2+4m_c^2\right)$.

Now, our calculation gives
\widetext
\begin{eqnarray}\label{eq:structf}
h_{\,^1\!S_0^{(8)}}\left(y,Q^2\right)&=&\left(g_V^c\right)^2 \,\times 6
\,{\frac{Q^2\, m_c^2} 
    {{\left(Q^2 + 4m_c^2 \right) }^2} }\,(y^2-2y+2)
\nonumber\\[.1in]
h_{\,^3\!S_1^{(8)}}\left(y,Q^2\right)&=&\left(g_A^c\right)^2 \,\times 2
m_c^2\,\frac{Q^2(y^2-2y+2)+16(1-y)m_c^2}{\left(Q^2+4m_c^2\right)^2}
\nonumber\\h_{\,^3\!P_0^{(8)}}\left(y,Q^2\right)&=& {\left(g_V^c\right)^2 \,\times 2\,{Q^2}\,}
      {\,\frac{     {{\left(Q^2 + 12m_c^2 \right) }^2}}{
     {{\left( {Q^2} + 4m_c^2  \right) }^4}}}\,(y^2-2y+2)
\\[0.1in]
h_{\,^3\!P_1^{(8)}}\left(y,Q^2\right)&=&\left(g_V^c\right)^2 \,\times 4\,Q^4\,{ \frac{
   Q^2(y^2-2y+2) + 16(1-y)m_c^2
     }{{{\left( {Q^2} + 4m_c^2 \right) }^4}}}
\nonumber\\h_{\,^3\!P_2^{(8)}}\left(y,Q^2\right)&=&
     \left(g_V^c\right)^2\,\times \frac{4}{5}\,Q^2\,{\frac{
      (y^2-2y+2){Q^4} +48(1-y)Q^2m_c^2+ 96(y^2-2y+2){{{m_c}}^4}
       }{ {{\left(Q^2 + 4m_c^2 \right) }^2}}}
   \nonumber
\end{eqnarray}
\narrowtext
where $g_A^c=\frac{1}{2}$ and $g_V^c=\frac{1}{2}\left(1-\frac{8}{3}\sin^2\Theta_W\right)$
are the vector and axial couplings of the $c$-quark. We have checked that an appropriately
modified version of these structure functions correctly reproduces the $e\, N\to e\, J/\psi\, X$
cross section in~Ref.~\cite{leptoprod}. We immediately observe that the coupling constants
favor the $\,^3S_1^{(8)}$ contribution, which is due to the large axial coupling (a similar
contribution is, of course, absent in the case of $J/\psi$ lepto and photoproduction). 
Indeed our numerical estimates will show that this matrix element dominates the total 
cross section, and also the differential
cross section unless $Q^2 \gg m_c^2$. At large $Q^2$, the relative $Q^4$ enhancement of the
$P$-wave structure functions makes them dominant.

These structure functions should eventually be incorporated in the 
specific Monte Carlo generators built for each particular detector.
Our numerical estimates presented in the following chapters ignore the 
fine details of the experiments and should be understood as
a preliminary feasibility study.

\section{Numerical results}\label{sec:numbers}

Due to its clean experimental signature, one of the most important charmonium 
production processes is $J/\psi$ production.
We have calculated the total cross section and the differential $Q^2$ distribution
of $J/\psi$'s at various experiments, both currently running and proposed. For lack
of knowledge of the size of the NRQCD matrix elements we set them equal to a reasonable 
``baseline size" value, defined~as
\widetext
\begin{equation}
\langle 0|{\cal O}_8\left(^1S_0\right)|0\rangle_{baseline}=
\langle 0|{\cal O}_8\left(^3S_1\right)|0\rangle_{baseline}=
\frac{\langle 0|{\cal O}_8\left(^3P_J\right)|0\rangle_{baseline}}{(2J+1)\,m_c^2}=
10^{-2}\,GeV^3
\end{equation}
\narrowtext
These values satisfy the restrictions imposed by heavy quark symmetry and are
compatible with the existing extractions~\cite{TeV1,B,Amud}. The results below were
found using MRST parton distribution functions (PDF)~\cite{MRST}, substituting 
$m_c=1.35\,GeV$ for the charm quark mass and imposing a cut $Q^2>(1.2\,GeV)^2$
in the calculation of the total cross section.

The energy dependence of the total cross section, taken at various representative incident
neutrino energies, is shown in Table~\ref{tab:crs-1}.
We find that the total cross section is very sensitive to the neutrino
beam energy $E_\nu$. The origin of this strong dependence can be traced to the fact
that the main contribution to the integral in~Eq.~(\ref{eq:struct1}) comes from
the smallest available $x\approx\frac{Q^2+4m_c^2}{s}$, so that larger $s$ probes
a smaller region of $x$ where the gluon PDF's are enhanced. It becomes obvious
that in order to have a reasonably large event number, the neutrino beams with the
highest possible energy should be employed.

\begin{table}[htb]
\begin{tabular}{ccccc}
$E_\nu[GeV]$   & 7.5  & 25    &  120  &   450  \\  \hline
$\sigma[nb]$     & $7.8\times10^{-13}$&  $6.9\times10^{-10}$ &
$1.3\times10^{-8}$ &  $5.5\times10^{-8} $
\end{tabular}
\caption{\label{tab:crs-1}Total cross sections for the $J/\psi$ production in
$\nu N\to J/\psi X$ for various representative incident neutrino energies.
The values are taken to correspond to the minimal, average and maximum neutrino beam 
energy at NOMAD and maximum available neutrino beam energy at NuTeV respectively.}
\end{table}
 
We have calculated the total cross section for the two presently running
high energy neutrino experiments, NOMAD at CERN and NuTeV at Fermilab, as well
as for the much-discussed high energy muon collider. Table~\ref{tab:crs} shows the
results for each term in the spectral decomposition. The values support our previous
expectations that the $^3S_1^{(8)}$ contribution dominates the total cross section. Note that these
numbers refer to direct $J/\psi$ production; a cascade mechanism involving
$\chi_{cJ}$ decay into $J/\psi$ can be easily calculated from our $\,^3S_1^{(8)}$
contribution and its ratio to direct $\,^3S_1^{(8)}$ production is universal and
can be taken from other experiments.

We should emphasize that due to the expected small number of events the 
question of background suppression becomes very important. In particular,
in the case of the currently running experiments, NOMAD and NuTeV, the
leptonic decay channel of $J/\psi$ becomes virtually the only possibility of
detecting the produced $J/\psi$'s. This makes electromagnetic lepton
pair production a very important source of background~\cite{em}.

\begin{center}
\begin{table}[htb]
\begin{tabular}{c||c|c|c|c|c|c|c|c}
{Spectral}&\multicolumn{2}{c|}{NOMAD}&
   \multicolumn{2}{c|}{NuTeV}&
   \multicolumn{2}{c|}{FMC-Ring}&
   \multicolumn{2}{c}{FMC-RLA3}\\
\cline{2-9}
decomp.\rule[-1.6ex]{0pt}{4.3ex}&$\sigma\left[10^{-9}\,nb\right]$&
   $10^6\frac{\sigma}{\sigma^{tot}}$&$\sigma\left[10^{-9}\,nb\right]$&
   $10^6\frac{\sigma}{\sigma^{tot}}$&$\sigma\left[10^{-9}\,nb\right]$&
   $10^6\frac{\sigma}{\sigma^{tot}}$&$\sigma\left[10^{-9}\,nb\right]$&
   $10^6\frac{\sigma}{\sigma^{tot}}$\\  \hline\hline
$^1S_0(8)$&$0.086$&$0.0131$&$1.59$&$1.94$&$2.01$&$3.89$&$1.54$&$2.30$\\
\tableline
$^3S_1(8)$&$0.552$&$0.0842$&$9.14$&$11.2$&$11.5$&$22.3$&$8.97$&$13.4$\\
\hline
$^3P_0(8)$&$0.129$&$0.0195$&$1.92$&$2.34$&$2.40$&$4.65$&$1.89$&$2.82$\\
\hline
$^3P_1(8)$&$0.082$&$0.0124$&$2.06$&$2.52$&$2.64$&$5.10$&$1.96$&$2.94$\\
\hline
$^3P_2(8)$&$0.166$&$0.0253$&$2.71$&$3.30$&$3.42$&$6.61$&$2.65$&$3.97$\\
\hline\hline
{\bf SUM}&$1.01$&$0.154$&$17.4$&$21.3$&$22.0$&$42.5$&$17.0$&$25.4$\\
\hline
{Total}&\multicolumn{2}{c|}{$\sigma^{tot}=6.56\,pb$}&
   \multicolumn{2}{c|}{$\sigma^{tot}=0.82\,pb$}&
   \multicolumn{2}{c|}{$\sigma^{tot}=0.52\,pb$}&
   \multicolumn{2}{c}{$\sigma^{tot}=0.67\,pb$}
\end{tabular}
\caption{\label{tab:crs}The $\nu N\to J/\psi X$ cross section contribution from each matrix element
(set equal to its baseline value) in the four discussed experiments, folding in the
neutrino energy distributions. Here, $\sigma^{tot}$ refers to the total deep inelastic cross 
section, $\sigma^{tot}=\sigma_{CC}+\sigma_{NC}$.}
\end{table}\end{center}

\begin{figure}[htb]
\begin{center}
\mbox{\epsfxsize=3in\epsfbox{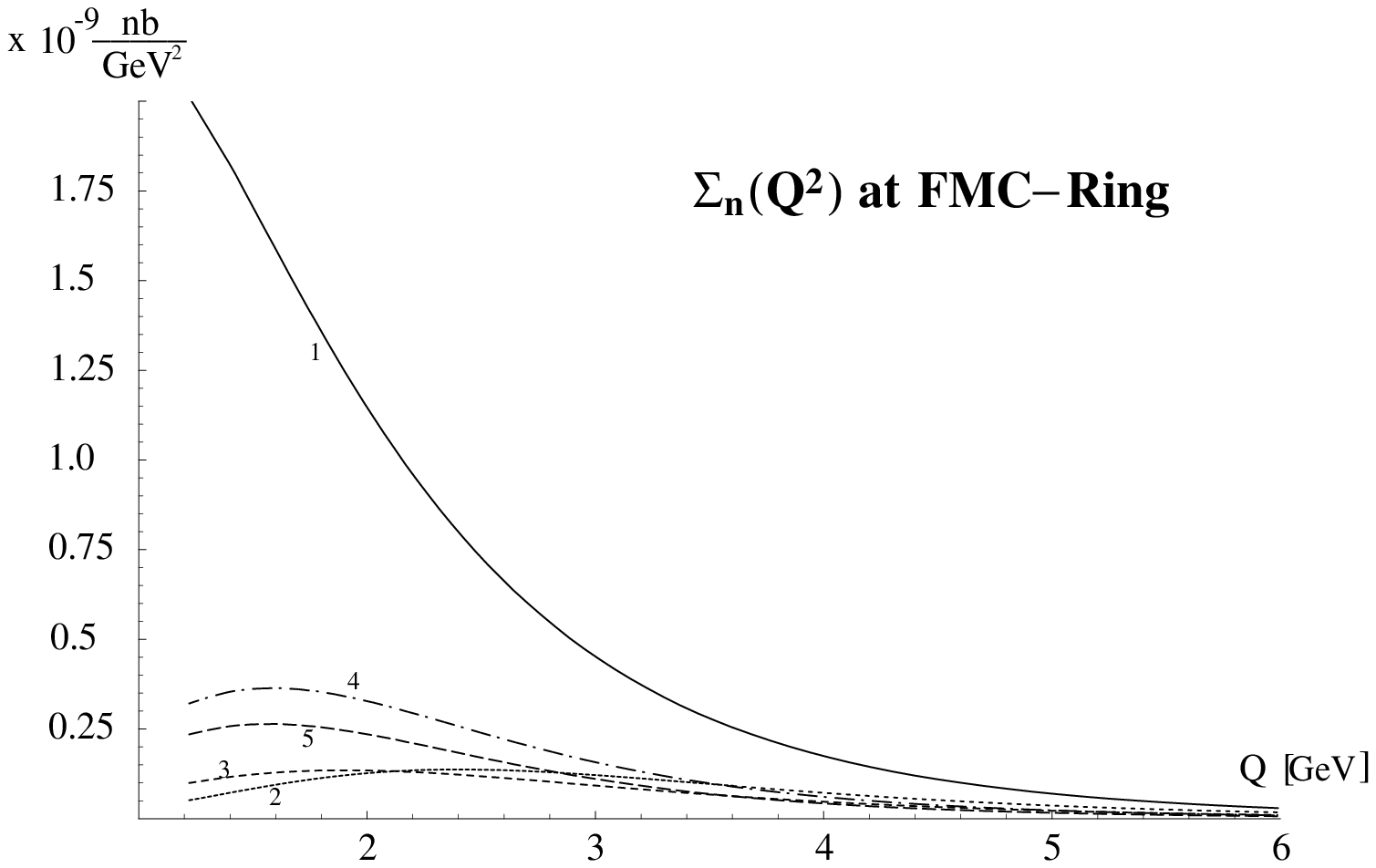}}
\mbox{\epsfxsize=3in\epsfbox{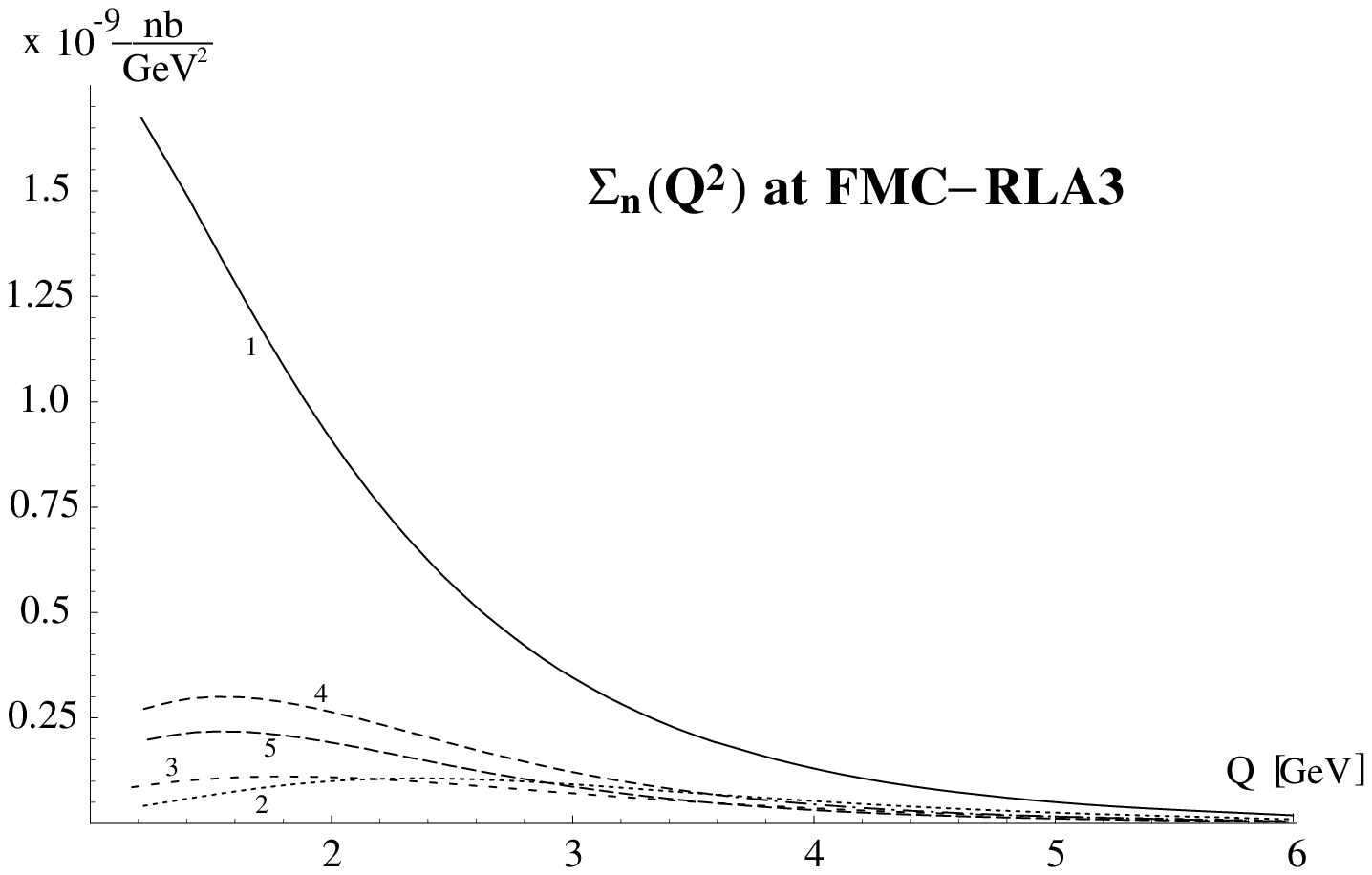}}
\mbox{\epsfxsize=3in \epsfbox{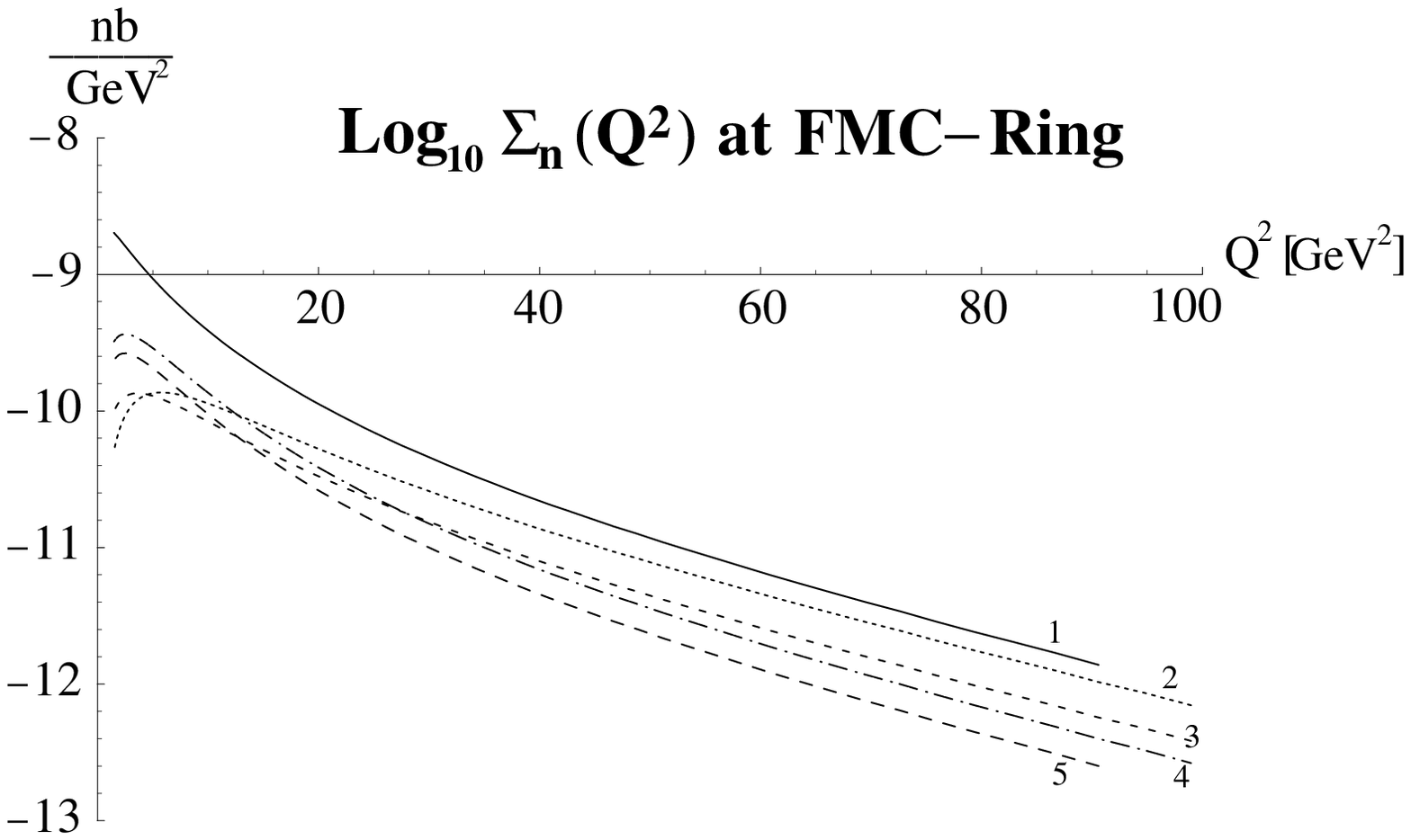}}
\mbox{\epsfxsize=3in \epsfbox{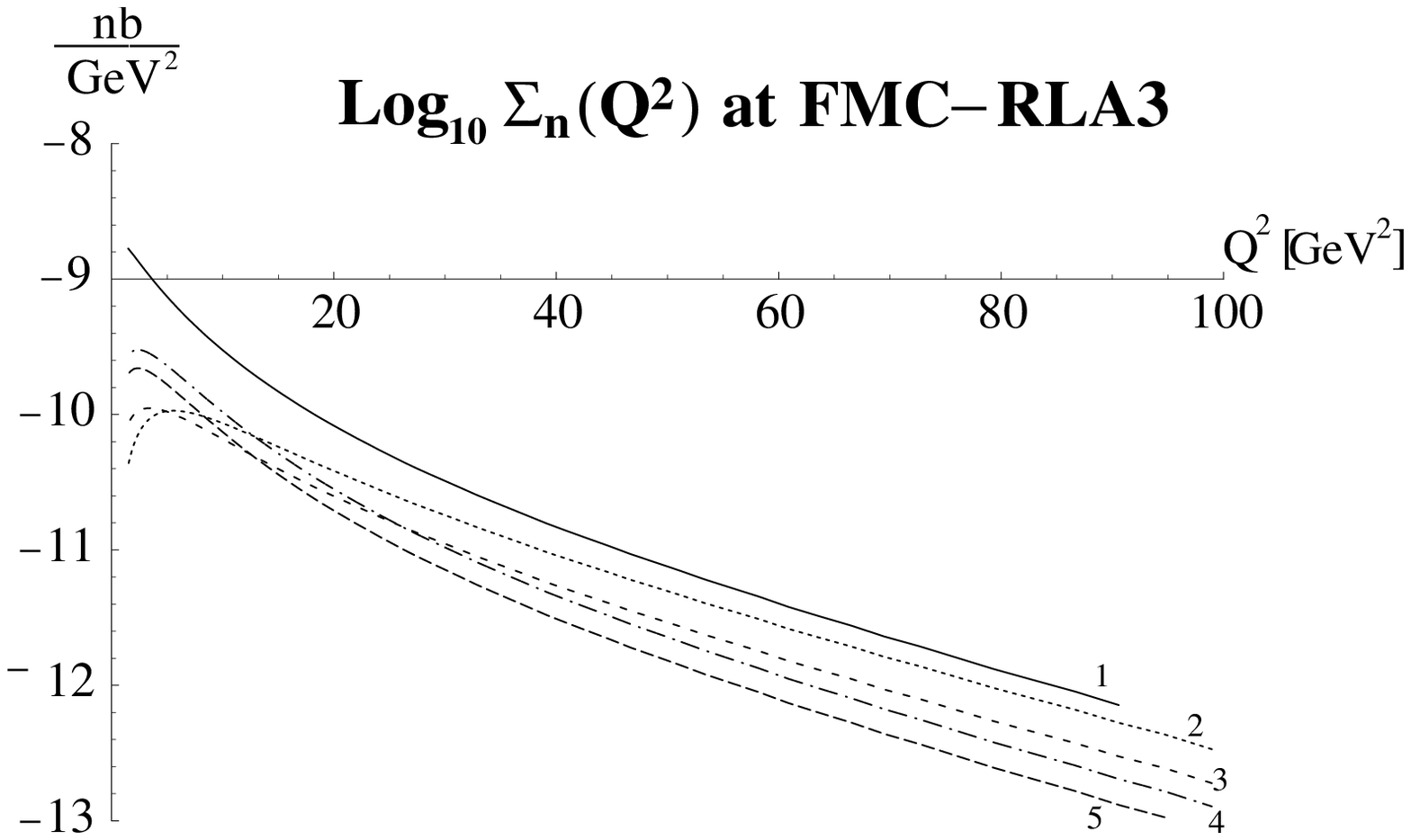}}
\end{center}
\caption{\label{FMC.Q3.il.eps}The $Q^2$ distribution
$\Sigma_n[\frac{nb}{GeV^2}]$ of the $J/\psi$ events at the FMC
for baseline-sized matrix elements: $d\sigma_n=\mu_nd\Sigma_n$
when $\langle{\cal O}_n\rangle=\mu_n\langle{\cal O}_n\rangle_{baseline}$.
The solid lines~(1) correspond to $\,^3S_1^{(8)}$, the dotted lines~(2) -- $\,^3P_1^{(8)}$,
the short-dashed lines~(3) -- $\,^1S_0^{(8)}$, the dash-dotted lines~(4) -- $\,^3P_2^{(8)}$,
and the long-dashed lines~(5) are the $\,^3P_0^{(8)}$ contributions.
}
\end{figure}

\begin{figure}[bt!]
\begin{center}
\mbox{\epsfxsize=3in\epsfbox{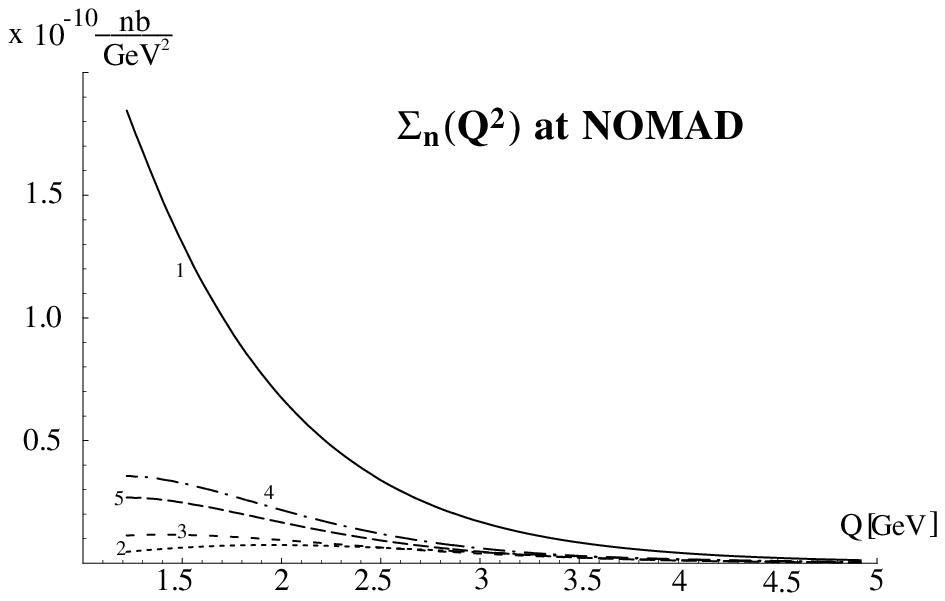}}
\mbox{\epsfxsize=3in\epsfbox{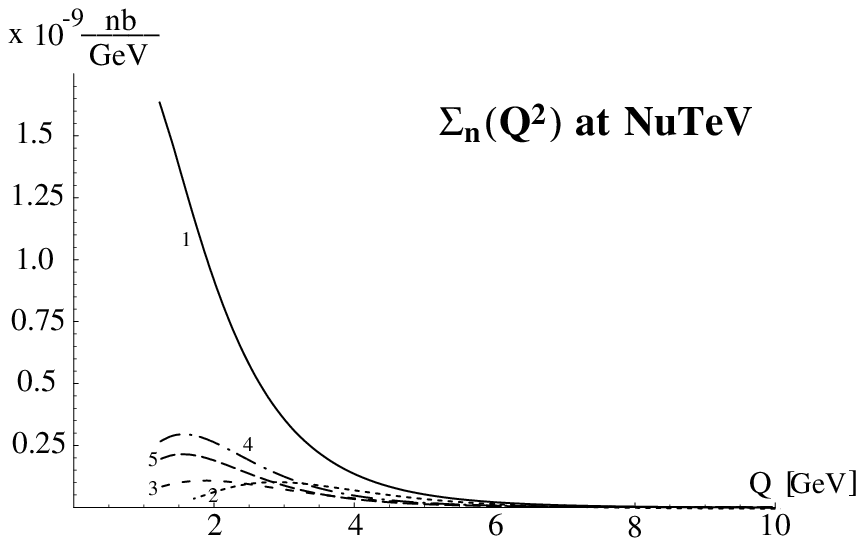}}
\caption{\label{NOMAD.Q3.il.eps}The $Q^2$ distribution
$\Sigma_n[\frac{nb}{GeV^2}]$ of the $J/\psi$ events at NOMAD and NuTeV.
The solid lines~(1) correspond to $\,^3S_1^{(8)}$, the dotted lines~(2) -- $\,^3P_1^{(8)}$,
the short-dashed lines~(3) -- $\,^1S_0^{(8)}$, the dash-dotted lines~(4) -- $\,^3P_2^{(8)}$,
and the long-dashed lines~(5) are the $\,^3P_0^{(8)}$ contributions.
}
\end{center}
\end{figure}

Another important issue is diffractive $J/\psi$ production.
It is difficult to separate, both theoretically and experimentally, diffractive
and color octet contributions. This is due to the fact that both of these processes contribute at 
$z \equiv (P_N\cdot P_\psi)/(P_N\cdot q) \sim 1$. In principle, the absence of
a rapidity gap in the color-octet NRQCD process may help~\cite{leptoprod}.
In addition, a perturbative QCD calculation of
the diffractive leptoproduction~\cite{Diffr} suggests that this process
falls off at high $Q^2$ as $\sim 1/Q^6$. We expect this behavior 
to hold for the case of diffractive ``neutrinoproduction'' as well.
As indicated in our calculation, the color octet contribution falls off
only as $\sim 1/Q^4$ which makes the diffractive process
negligible at sufficiently high $Q^2$.
Figs. \mbox{\ref{FMC.Q3.il.eps} and \ref{NOMAD.Q3.il.eps}}
show the $Q^2$ distribution of the events. As expected, these decrease
quickly with increasing~$Q^2$. The decrease, however, is not so drastic as not
to let us use higher $Q^2$ cut up to ${\cal O}(10\,GeV^2)$ which may
also allow to reduce the nonperturbative background.

Now we turn to the discussion of some of the presently running and
planned neutrino experiments.

\subsection{The muon collider}
A proposal to build a high energy $\mu^+\mu^-$ storage ring has recently
drawn much attention. Although seemingly a sideline, luminous neutrino beams
are unavoidably generated by the decaying muons, and provide an
ideal environment for charmonium generation. Such machines comprise both
necessary ingredients, high luminosity and high neutrino beam energy, that
are needed for detailed studies of charmonium production.

The proposed First Muon Collider~\cite{Quigg} would use $2\times250\,\rm{GeV}$ muons,
boosted in linear accelerators. These muons and antimuons decay into the $\nu_\mu$'s 
(or $\overline{\nu}_\mu$'s) in both final booster (called RLA3) and in the 
accelerator ring. A straight section of the ring
was proposed in order to achieve a highly collimated neutrino beam.
The resulting neutrino spectra that have been calculated by Harris and
McFarland~\cite{Harris2} peak in the $150-200\,\rm{GeV}$ region. With a representative
set of parameters ${\cal O}(10^6)$ DIS events/(g/cm$^2$ year) are expected. 
This high number of events allows a measurement of the partial
distributions without the need for extremely costly dedicated detectors. With
the cross sections given in Table~\ref{tab:crs}, this translates to
$20-50~J/\psi$ events/(g/cm$^2$ year).

We will now sample the various suggestions for the neutrino detectors to get a rough
idea of the yield and capabilities. In all cases we look at a detector diameter
of $40\,\rm{cm}$, in which almost all of the neutrino beam fits.

As an obvious minimal suggestion, we might look at a $30\,\rm{cm}$ thick, table-top sized
detector~\cite{Harris2}, containing approximately $50\,\rm{kg}$ of water-density target.
We estimate a yield of $1600~J/\psi\mbox{~events}/\mbox{~year}$, providing just
enough statistics for a crude estimate of the octet matrix elements. By contrast,
a light target considered e.g. in~\cite{Harris1} (a $1\,\rm{m}$ thick liquid
hydrogen detector) would only produce approximately $100~J/\psi\mbox{~events}/\mbox{year}$,
not much better than presently running experiments.

The best answer may be the ``general purpose detector" suggested by B.~King~\cite{BKing}.
This detector consists of a one meter long stack of silicon CCD tracking planes,
providing $\sim50$ g/$\mbox{cm}^2$ density. In such a detector an estimated
$3,000~J/\psi\mbox{~events}/\mbox{year}$ would occur and the high efficiency and precise
reconstruction capabilities should allow for disentangling the contribution from
the color octet $\,^3S_1^{(8)}$ and $\,^3P_J^{(8)}$ matrix elements, through the measurement
of the differential $Q^2$ distributions. Note that an increase in the length of the
straight section of the storage ring provides a relatively cheap way of increasing the
luminosity of the neutrino beam, resulting in better statistics. Ref.~\cite{BKing}
actually considers a $200\,\rm{m}$ long straight section, gaining a factor of $20$ over
our estimates.

Another option is a conventional fixed target type heavy detector~\cite{Harris2}.
A representative example of a 2 ton iron calorimeter $1.6\,\rm{m}$ long, would see 
very high event rates (approximately
$70,000~J/\psi\mbox{~events}/\mbox{year}$ before cuts and detection
efficiencies are included). If background difficulties can be overcome, this
option provides an excellent opportunity to study all differential rate distributions
with precise statistics.

As Fig.~\ref{FMC.Q3.il.eps} shows, the differential cross section quickly decreases with $Q^2$.
However, the decrease is not as fast as in the leptoproduction case. Comparing the $Q^2$
behavior of formula for the $\,^3S_1^{(8)}$ contribution to Eq.~(3) in 
Ref.~\cite{leptoprod}, we find
\begin{equation}
\left(\frac{d\sigma}{dQ^2}\right)_{\nu N} \propto \left(\frac{d\sigma}{dQ^2}\right)_{eN}
\times \frac{Q^2(Q^2 + \beta m_c^2)}{m_Z^4}
\end{equation}
where $\beta \sim 4-10$ is slightly $Q^2$-dependent. We have checked that our curves
actually reproduce this relationship with $\beta \approx6$. The upshot is that the absence of
a photon propagator results in a much wider tail of the $Q^2$ distribution than
in leptoproduction, allowing for better discrimination between
$S$- and $P$-wave contributions.
\begin{figure}[tbh]
\begin{center}
\mbox{\epsfxsize=3in \epsfbox{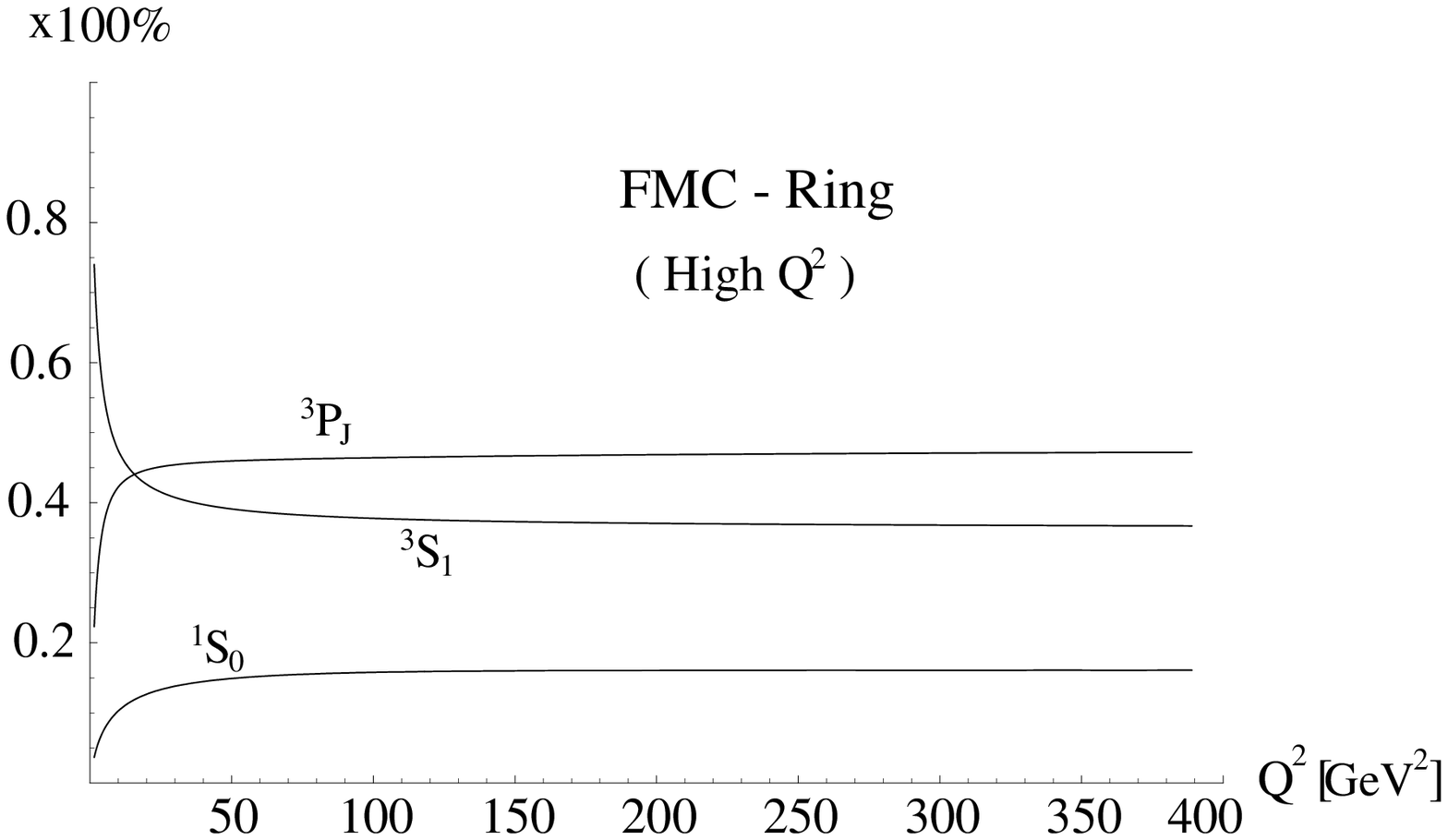}}
\mbox{\epsfxsize=3in \epsfbox{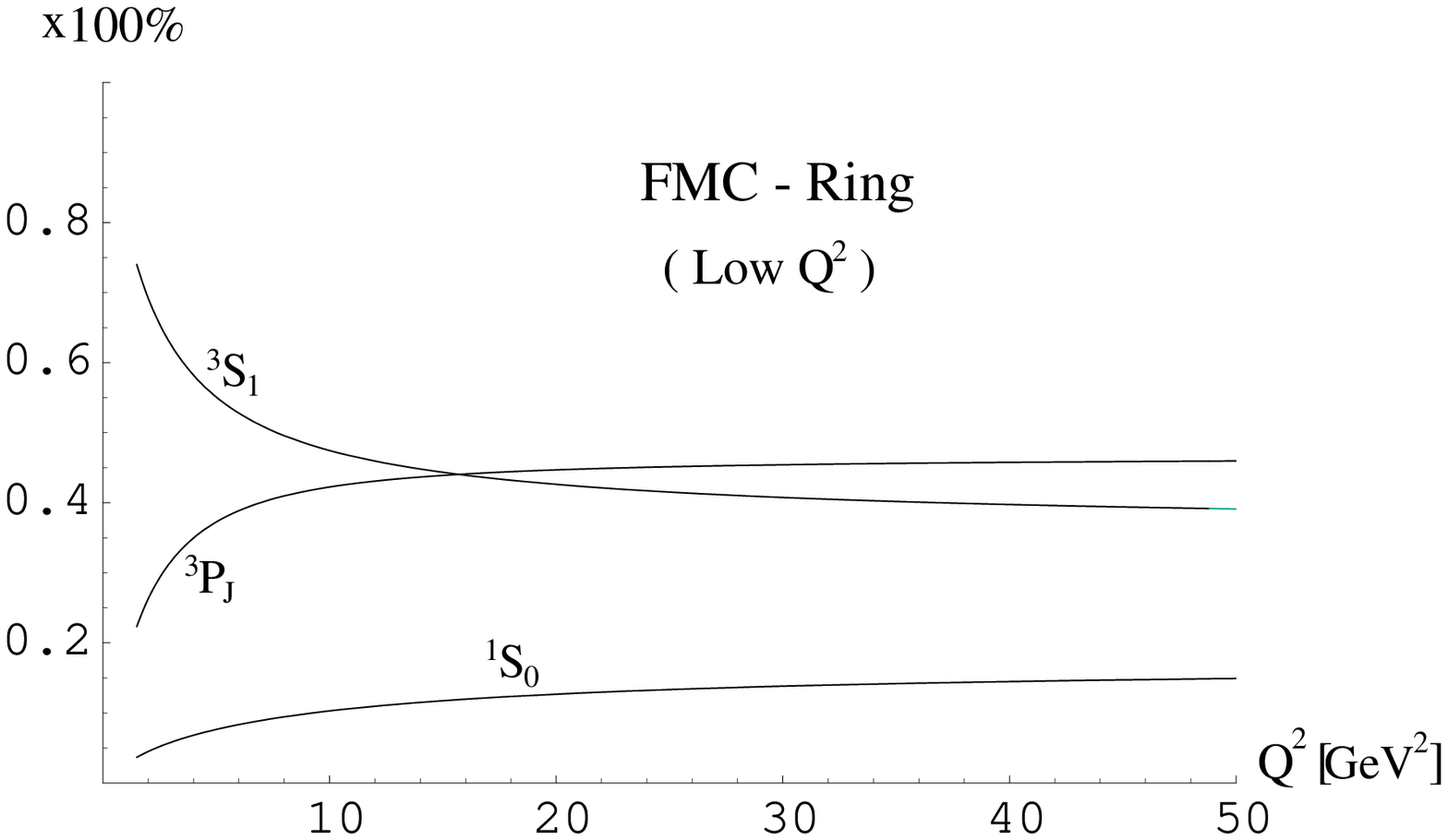}}
\end{center}
\caption{\label{FMC-Ring.ratio.1.eps}The contributions of the various matrix
elements as a function of $Q^2$ for FMC-Ring, with baseline-sized matrix elements.
Observe that the contributing linear combination of matrix elements drastically changes
with $Q^2$ within a region where the rate is still sizable, so one can realistically
extract both the $\,^3S_1^{(8)}$ and the $\,^3P_J^{(8)}$ matrix elements even without
requiring exceedingly high statistics.}
\end{figure}

Fig.~\ref{FMC-Ring.ratio.1.eps} shows the ratio of the contributions from these matrix elements
as a function of $Q^2$. At low $Q^2$, where the contribution into the cross section is
large, the $\,^3S_1^{(8)}$ dominates but a determination of the differential cross section
in the range $\sim2\,GeV<Q<4\,GeV$, where the event number is still sizable, would allow
a reliable extraction of both $\,^3S_1^{(8)}$ and $\,^3P_J^{(8)}$ matrix elements. Higher
$Q^2$'s will only measure a linear combination of these, as the ratio in
Fig.~\ref{FMC-Ring.ratio.1.eps} becomes $Q^2$-independent.

Given the above calculation, it is easy to estimate the production rate of other
charmonium states, for instance $\eta_c$. Using heavy quark symmetry, it is possible
to relate, at leading order in $v^2$, the leading ${\cal O}(v^7)$ matrix elements
of $\eta_c$, $\,^1S_0^{(8)},\,^3S_1^{(8)},\,^1P_1^{(8)}$ to the corresponding $J/\psi$ matrix
elements. A simple calculation shows, using the data in Table~\ref{tab:crs} with baseline-sized
matrix elements, that the $\eta_c$ yield will be one third of the $J/\psi$'s. This fact and
the absence of a clear signature will make $\eta_c$ quite elusive in this experiment.
A similar remark holds for the $h_c$, whose only contributing matrix element, the
${\cal O}(v^5)$ level $\,^1S_0^{(8)}$, is related to the $\,^3S_1^{(8)}$ matrix elements of
the $\chi_{cJ}$'s.

\subsection{NOMAD}

The NOMAD detector at CERN uses the CERN SPS neutrino beam, mainly designed for
detecting neutrino oscillations. The average neutrino energy is small, $24\,\rm{GeV}$,
and, even with the huge mass of the detector, the event number is small. One would
expect approximately $2-3$ events per year in the main detector with its fiducial mass of $2.7$ tons.
The situation is somewhat better in the Front Calorimeter, where such a search is underway, simply
because of the larger mass of the detector.\footnote{The authors thank
Kai Zuber for drawing their attention to this point.} With a mass of $17.7$ tons,
and the detector characteristics given in Ref.~\cite{nomadpaper}, the yield
becomes $14-15~J/\psi\mbox{~events}/\mbox{year}$, which must be multiplied by the decay ratios and
detector efficiencies to find the number of observable events. Given the inaccuracies
of our calculation and the crude estimate of the size of the matrix element, it does
not seem impossible that several such events would be seen.

\subsection{NuTeV}
Another high energy neutrino oscillation experiment has a chance of probing, as
a byproduct, our process. The larger average neutrino  energy at the Fermilab
experiment, compared to CERN, results in approximately twenty times larger cross section,
compensating for the smaller size of the detector. The Fermilab experiment NuTeV
detected, during its lifetime of one year, about $1.3$ million DIS events. Using
the numbers in Table~\ref{tab:crs}, this would imply $28\,J/\psi$ events. This is
again on the verge of observability.

\section{Conclusions}\label{sec:concl}

In conclusion we note that an analogous calculation can be done for the $b\overline b$ system.
In that case, however, the $\alpha_s$ suppression of the ratio of the color singlet vs. color octet
contributions is compensated by the smallness of the heavy quark velocity. Indeed, we expect
\widetext
\begin{equation}
\frac{(singlet/octet)_{b\overline b}}{(singlet/octet)_{c\overline c}}\sim
\frac{\left(\alpha_s/v^4\right)_{b\overline b}}{\left(\alpha_s/v^4\right)_{c\overline c}}\sim6
\ \ \mbox{because}\ \ \frac{\alpha_s(m_\Upsilon)}{\alpha_s(m_{J/\psi})}\sim\frac{2}{3},\ \ \mbox{and}
\ \ \left(\frac{v_c^2}{v_b^2}\right)^2\sim\left(\frac{0.3}{0.1}\right)^2\sim9.
\end{equation}
\narrowtext
(Recall that in $J/\psi$ leptoproduction the color singlet contribution was suppressed
by~$\sim\,10\%$.) As a consequence, our tree level calculation is unable to accurately predict 
the dominant $\Upsilon$ production rate. However, because all numerical factors in the color octet 
contribution are smaller for the $\Upsilon$ than for the $J/\psi$ (including the size of the
phase space), we can at least exclude the possibility of $\Upsilon$ production at presently
running experiments. In order to make a meaningful estimate for the muon collider, one needs
to perform much more involved ${\cal O}(\alpha_s^2)$ calculations.

To summarize, we have investigated charmonium production at some of the currently running
(NOMAD and NuTeV) and proposed (First Muon Collider) neutrino factories.
We found that the $J/\psi$ production cross section is dominated by the contribution from 
the matrix element of the color octet $^3S_1^{(8)}$ operator, which is not
accessible in photo and leptoproduction experiments.
Our exploratory study shows that the neutrino beam from the $\mu^+\mu^-$ collider 
can be successfully used to accurately extract the color octet $\,^3S_1^{(8)}$ and 
$\,^3P_J^{(8)}$ matrix elements of the $J/\psi$. This
measurement is complementary to leptoproduction, where $\,^3S_0$ and $\,^3P_J$ contribute.
In order to achieve an accurate extraction, however, a detailed investigation is necessary,
including the ${\cal O}(\alpha_s^2)$ contributions. A significant deviation of the
measured values of the nonperturbative charmonium matrix elements extracted from
high energy $\nu p$ and $p \bar p$ experiments could be a sign of intrinsic charm in 
the proton.

\section*{Acknowledgments}

We would like to thank Sean Fleming, Tom Mehen and Adam Falk for useful discussions. 
We also acknowledge the help of the experimentalists Kai Zuber at NOMAD and 
Todd Adams and Drew Alton at NuTeV for answering our questions concerning their 
experiments. This work was supported in part by 
the United States National Science  Foundation under Grant
No.~PHY-9404057 and No.~PHY-9457916, by the United States Department of 
Energy under Grant No.~DE-FG02-94ER40869, and by a grant from the NSERC of Canada.


\thebibliography{99}

\bibitem{NRQCDrev} G.T.~Bodwin, E.~Braaten and G.P.~Lepage,
 Phys. Rev. {\bf D51}, 1125 (1995), hep-ph/9407339;
For a recent review see, for example, 
B.~Grinstein, ``A Modern introduction to quarkonium theory,"
hep-ph/9811264.

\bibitem{Bc} E.~Braaten and S.~Fleming,
Phys. Rev. {\bf D52}, 181 (1995), hep-ph/9501296;
G.~Chiladze, A.F.~Falk and A.A.~Petrov,
hep-ph/9811405.

\bibitem{Hh} A.A.~Petrov,
{\it To be published in the proceedings of 3rd International Conference in 
Quark Confinement and Hadron Spectrum (Confinement III), Newport News, 
VA, 7-12 Jun 1998}, hep-ph/9808347; 
G.~Chiladze, A.F.~Falk and A.A.~Petrov,
Phys. Rev. {\bf D58}, 034013 (1998), hep-ph/9804248.

\bibitem{TeV1} P.~Cho and A.K.~Leibovich,
Phys. Rev. {\bf D53}, 150 (1996), hep-ph/9505329;
{\bf ibid} {\bf D53}, 6203 (1996), hep-ph/9511315; 
M.~Beneke and M.~Kramer,
Phys. Rev. {\bf D55}, 5269 (1997), hep-ph/9611218.

\bibitem{leiblep} C.G.~Boyd, A.K.~Leibovich and I.Z.~Rothstein,
Phys. Rev. {\bf D59}, 054016 (1999), hep-ph/9810364. 

\bibitem{B} M.~Beneke, F.~Maltoni and I.Z.~Rothstein,
Phys. Rev. {\bf D59}, 054003 (1999), hep-ph/9808360.

\bibitem{leptoprod} S.~Fleming and T.~Mehen,
Phys. Rev. {\bf D57}, 1846 (1998), hep-ph/9707365.

\bibitem{Amud}J.~Amundson, S.~Fleming and I.~Maksymyk,
Phys. Rev. {\bf D56}, 5844 (1997), hep-ph/9601298.

\bibitem{MRST}A.D.~Martin, R.G.~Roberts, W.J.~Stirling and R.S.~Thorne,
Eur. Phys. J. {\bf C4}, 463 (1998), hep-ph/9803445 and
references therein.

\bibitem{em} R.~Belusevic and J.~Smith,
Phys. Rev. {\bf D37}, 2419 (1988).

\bibitem{Diffr} M.G.~Ryskin, R.G.~Roberts, A.D.~Martin and E.M.~Levin,
Z. Phys. {\bf C76}, 231 (1997), hep-ph/9511228;
S.J.~Brodsky, L.~Frankfurt, J.F.~Gunion, A.H.~Mueller and M.~Strikman,
Phys. Rev. {\bf D50}, 3134 (1994), hep-ph/9402283.

\bibitem{Quigg} C.~Quigg, hep-ph/9803326.

\bibitem{Harris2} D.A.~Harris and K.S.~McFarland, hep-ex/9804009.

\bibitem{Harris1} D.A.~Harris and K.S.~McFarland, hep-ex/9804010.

\bibitem{BKing} B.J.~King,
{\it  In *Batavia 1997, Physics at the first muon collider* 334-348}.

\bibitem{nomadpaper} J.~Altegoer {\it et al.} [NOMAD Collaboration],
Nucl. Instrum. Meth. {\bf A404}, 96 (1998).
%
\end{document}